\documentstyle[epsfig]{aipproc}

\newcommand{\kpnn}    {\mbox{$K^+ \! \rightarrow \! \pi^+ \nu \overline{\nu}$ }}
\newcommand{\kzpnn}   {\mbox{$K \! \rightarrow \! \pi \nu \overline{\nu}$ }}
\newcommand{\kzppnn}  {\mbox{$K \! \rightarrow \! \pi \pi \nu \overline{\nu}$ }}
\newcommand{\klpnn}   {\mbox{$K^\circ_L \! \rightarrow \! \pi^\circ \nu \overline{\nu}$ }}
\newcommand{\kppnn}   {\mbox{$K^+ \! \rightarrow \! \pi^+ \pi^\circ \nu \overline{\nu}$ }}
\newcommand{\kmng}    {\mbox{$K^+ \! \rightarrow \! \mu^+ \nu_\mu \gamma$ }}
\newcommand{\kppg}    {\mbox{$K^+ \! \rightarrow \! \pi^+ \pi^\circ \gamma$ }}

\newcommand{\kzpen}   {\mbox{$K\!\rightarrow\!\pi e \nu_e$ }}
\newcommand{\vtd}     {\mbox{$V_{td}$ }}
\newcommand{\Vtd}     {\mbox{$| V_{td} |$ }}
\newcommand{\bpsiks}  {\mbox{$B^\circ_d \! \rightarrow \! \psi K^\circ_S$ }}

\begin{document}
\title{ 
\vspace{-1.5cm}
\rightline{\small\rm BNL--67630}
\vspace{-0.3cm}
\rightline{\small\rm August 15, 2000}
\vspace{1cm}
Measurement of B(\kpnn)\footnote{ To be published in the 
{\it Proceedings of the  7$^{th}$ Conference on the Intersections 
of Particle and Nuclear Physics;
Quebec City, Canada, May 22-28, 2000};
Z.~Parsa and W.~Marciano, Eds.  }
 }

\author{S.H.~Kettell\footnote{for the E787 collaboration}}
\address{Brookhaven National Laboratory}

\maketitle

\begin{abstract}
The experimental measurement of \kpnn is reviewed.  New results from
experiment E787 at BNL are presented: with data from 1995--97 the
branching ratio has been measured to be B(\kpnn) =
$(1.5^{+3.4}_{-1.2}) \times10^{-10}$. The future prospects for
additional data in this mode are examined.
\end{abstract}

\section*{Introduction}

The unprecedented sensitivities of rare kaon decay experiments and the
recent observation of \kpnn have opened doors to the measurement of
the unitarity triangle completely within the kaon system.  The decay
\kpnn is one of the `golden modes' for measuring CKM
parameters. Measurement of the branching ratio B(\kpnn) provides a
clean and unambiguous determination of the CKM matrix element \Vtd, in
particular of the quantity $|\lambda_t|\equiv |V^*_{ts}V_{td}|.$

The theoretical uncertainty in \kpnn is quite small ($\sim$7\%) as the
hadronic matrix element is extracted from the \kzpen branching ratio
B($K_{e3}$).  The \kpnn branching ratio has been calculated to
next-to-leading-log approximation~\cite{bb1}, with isospin violation
corrections~\cite{marciano} and two-loop-electroweak
effects~\cite{bb2}.  The branching ratio can be expressed
as~\cite{bb3}
\begin{eqnarray}
B(\kpnn) & = &\frac{\kappa_+\alpha^2 B(K_{e3})}
        {2\pi^2 \sin^4 \theta_W |V_{us}|^2}
      \sum_l  | X_t\lambda_t + X_c\lambda_c |^2 \\ \nonumber
& = & 8.88 \times 10^{-11} A^4
[(\overline{\rho}_0-\overline{\rho})^2 + (\sigma\overline{\eta})^2]
\\ \nonumber
& = & 3.6\times10^{-4}([Re(\lambda_t)-1.4\times10^{-4}]^2 +
[Im(\lambda_t)]^2) \\ \nonumber
         & = & (8.2\pm3.2)\times10^{-11},
\end{eqnarray}
with the error determined by the present uncertainty of $\lambda_t$.
The current limit on $B_s$---$\overline{B}_s$ mixing, combined with
the measured frequency of $B_d$---$\overline{B}_d$ mixing, implies a limit
on \kpnn that has very small theoretical uncertainty~\cite{bb3}:
B(\kpnn) $ < 1.67\times10^{-10}$.

\section*{Experiment E787}

The E787 experiment at BNL~\cite{e787_det} was designed to search for
\kpnn and collected data in 1989--91\cite{e787_pnnold} and then again
in 1995--98 after an upgrade to the detector and
beamline~\cite{e787_det2}.

\begin{figure}[h] 
 \begin{minipage}{0.35\linewidth}
  {\large a)}
  \epsfig{file=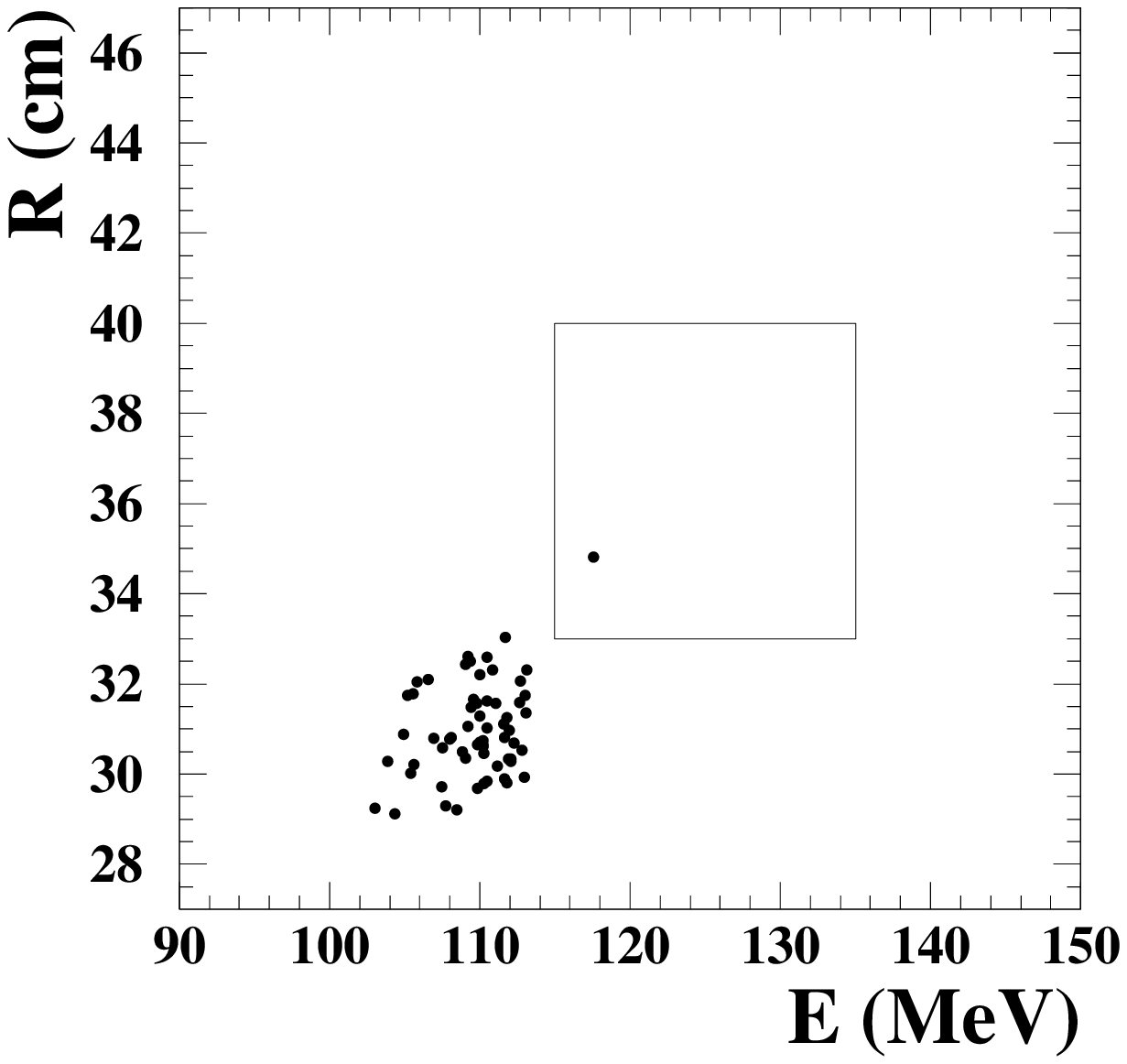,height=2.5in,width=2.5in}
 \end{minipage}\hfill
 \begin{minipage}{0.55\linewidth}
  {\large b)}
  \epsfig{file=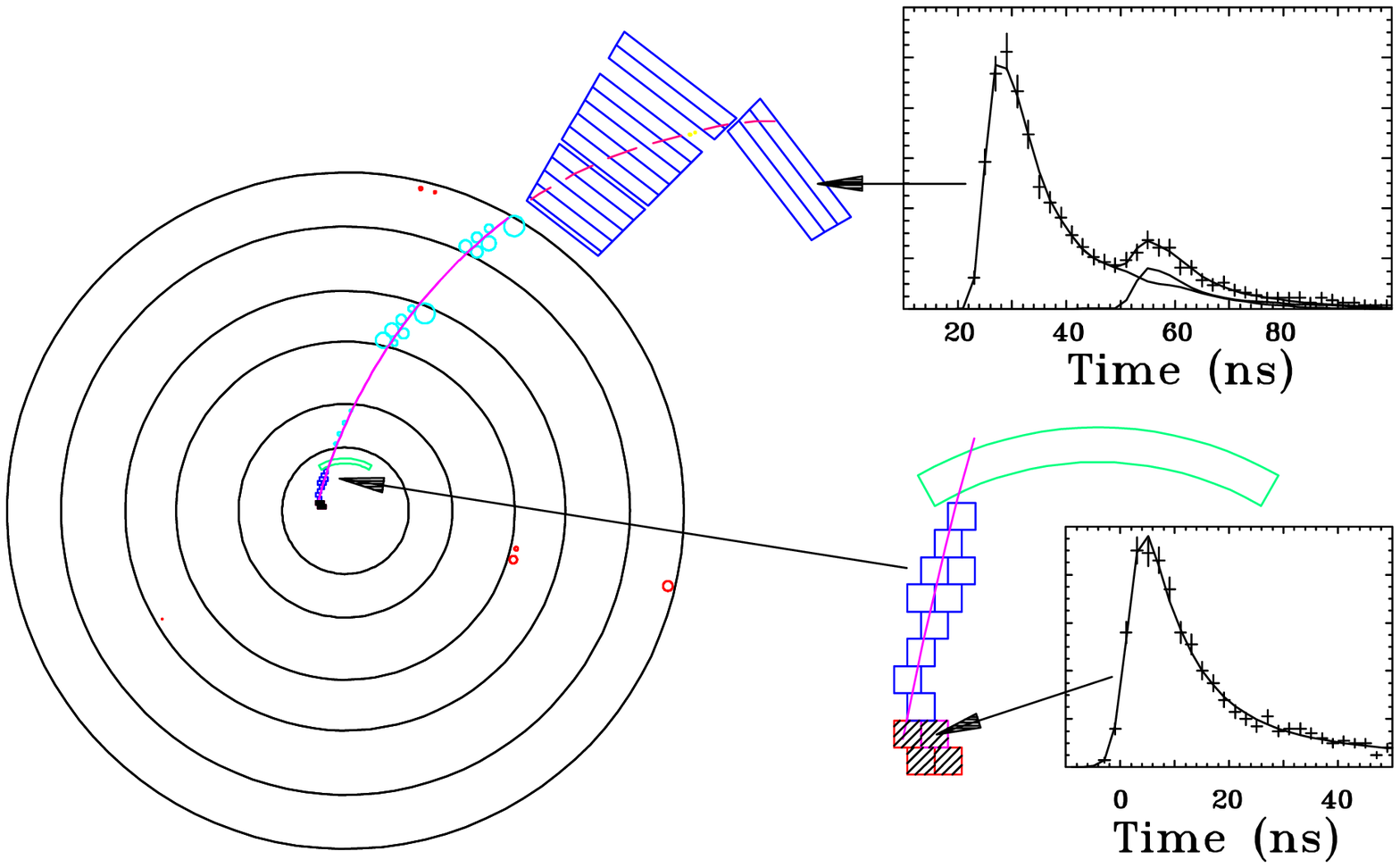,height=2.in,width=3.in}
 \end{minipage}\hfill
\caption{(a) Range vs. Kinetic energy plot of the final sample.  The
events near $E=108$~MeV are $K_{\pi 2}$ background. The box indicates
the accepted region for \kpnn events.  (b) Event display of the \kpnn
signal event.}
\label{fig:r_e}
\end{figure}

The first observation of \kpnn was reported in the analysis of the
1995 data set~\cite{pnn95}.  A new and improved analysis of the
1995--97 data set has reduced the background levels by almost a factor
of three.  A plot of range vs. energy for events passing all other
\kpnn criteria is shown in Figure~\ref{fig:r_e}(a).
One event is observed in the signal region (the same one as observed
in the 1995 data: see Ref.~\ref{ref:pnn95}).  A reconstruction of the
event is shown in Figure~\ref{fig:r_e}(b).  The background is measured
to be $0.08\pm0.02$ events.  Based on this one event, the branching
ratio is B(\kpnn) = $(1.5^{+3.4}_{-1.2}) \times
10^{-10}$~\cite{e787_pnn}. From this measurement, a limit of
\begin{equation}
0.002 < |\vtd| < 0.04
\end{equation}
is determined; in addition, the following limits on $\lambda_t \equiv
V^*_{ts}\vtd$ can be set:
\begin{eqnarray}
& | Im(\lambda_t) |  &  <   1.22\times10^{-3} \\ \nonumber
-1.10\times10^{-3} < & Re(\lambda_t) & < 1.39\times10^{-3} \\ \nonumber
1.07\times10^{-4}  < & | \lambda_t | & < 1.39\times10^{-3}.
\end{eqnarray}
The final sensitivity of the E787 experiment, based on data from
1995--98, should reach a factor of two further to the standard-model
(SM) expectation for \kpnn.

The 90\% CL upper limit on \kpnn is B(\kpnn) $ < 5.8\times 10^{-10}$
and a model-independent limit~\cite{grossman} on the neutral mode,
\klpnn, can be derived from this result:
\begin{eqnarray}
B(\klpnn) & < & 4.4 \times  B(\kpnn) \\ \nonumber
          & < &  2.6\times10^{-9} \; \; (90\%\,{\rm CL})
\end{eqnarray}

The \kzppnn decays, such as \kppnn, can also, in principle, provide a
clean determination of CKM matrix parameters; however, due to the
small SM branching ratios, e.g. B(\kppnn) = (1--2)$\times10^{-14}$,
their usefulness is limited.  E787 has recently set the first limit on
any of these modes~\cite{e787_ppnn}, with B(\kppnn) $<
4.3\times10^{-5}$.  Additional new results from E787 include a
measurement of the direct emission (DE) radiation in \kppg decay
(K$_{\pi2\gamma}$) and the first observation of structure-dependent
radiation (SD) in \kmng decays (K$_{\mu2\gamma}$).  With eight times
higher statistical sensitivity than previous experiments and better
kinematic constraints E787 has measured a branching ratio for direct
emission B(K$_{\pi2\gamma}$:DE) that is four times smaller than
previous results~\cite{e787_ppg}, B(K$_{\pi2\gamma}$:DE, 55 MeV $<$
$T_+$ $<$ 90 MeV) = $(4.7\pm0.8\pm0.3)\times 10^{-6}$ ($T_+$ is the
kinetic energy of the $\pi^+$).  E787 has also measured the branching
ratio for structure-dependent K$_{\mu2\gamma}$~\cite{e787_mng} to be
B(K$_{\mu2\gamma}:SD^+$) = $(1.33 \pm 0.12 \pm 0.18) \times
10^{-5}$. The vector and axial-vector form factors are $|F_V+F_A|
=0.165 \pm 0.007 \pm 0.011$ and $-0.04 < F_V-F_A < 0.24$ at $90\%$ CL.

\section*{Future Prospects}

Significant progress in the determination of the fundamental CKM
parameters from the \kzpnn system will be made in the generation of
experiments that is now starting: E949 and KOPIO at BNL, CKM and KAMI
at FNAL and E391 at KEK.  These measurements can unambiguously
determine SM {\it CP} violation parameters. Comparison with the
B-system will then over-constrain the triangle and test the SM
explanation of {\it CP} violation.  The two most important tests are
expected to be:
\begin{itemize}
  \item Comparison of the angle $2\beta$ from the ratio
B(\kpnn)/B(\klpnn) and the {\it CP} asymmetry in the decay
\bpsiks~\cite{grossman,sinb}.
  \item Comparison of the magnitude $|\vtd|$ from \kpnn and the ratio of
the mixing frequencies of $B_s$ and $B_d$ mesons~\cite{bb3}.
\end{itemize}
Improvement in the charged mode \kpnn will come in two steps: E949 and
CKM.

The E787 experiment has already demonstrated sufficient background
rejection for a very precise measurement of B(\kpnn). A new experiment
under construction, E949, is expected to run in 2001--03. Taking
advantage of the very large AGS proton flux and the experience gained
with the E787 detector, E949 with modest upgrades should observe $\cal
O$(10) SM events in a two year run. The background is well-understood
and is $\sim$10\% of the SM signal.

A proposal for a further factor of 10 improvement has been initiated
at FNAL. The CKM experiment (E905) plans to collect $\cal O$(100) SM events,
with $\cal O$(10) background events, in a two year run starting after
2005.  This experiment will use a new technique, with K$^+$
decay-in-flight and independent momentum (Si and straw-tube trackers) and
velocity (kaon and pion RICH) spectrometers. CKM is situated in a high
flux 22 GeV/c RF-separated kaon beam, derived from the Main Injector
at FNAL.

\begin{figure}[ht]
\centerline{\epsfig{file=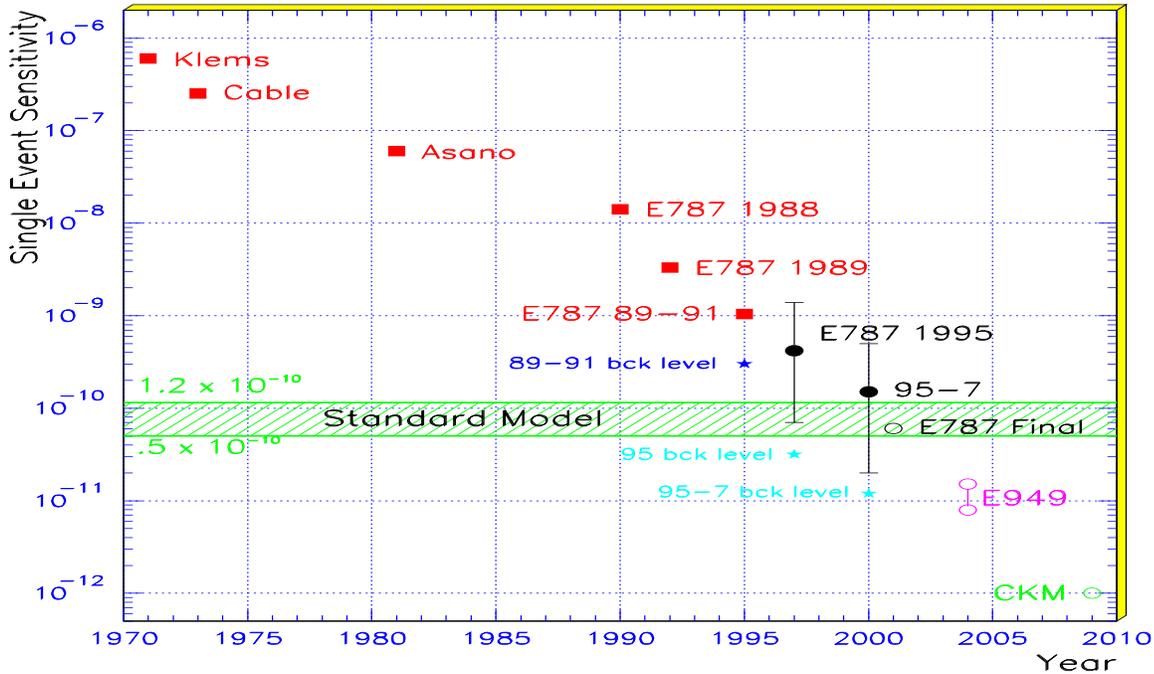,height=4.in,width=6.5in}}
\caption{Sensitivities of experiments searching for \kpnn. The
sensitivities of experiments setting limits are shown as red
squares. The projected sensitivities are shown as open
circles. Background measurements for several of the recent E787
searches are shown as blue stars.}
\label{fig:hist}
\end{figure}

Improvements in the sensitivities of \kpnn experiments over time is
shown in Figure~\ref{fig:hist}.
Published data are shown as solid points and future projections 
to E949 and CKM are shown as open points.
 
\section*{Acknowledgments}

This work was supported under U.S. Department of Energy contract
\#DE-AC02-98CH10886.

\def\Journal#1#2#3#4{{#1}{\bf #2}, #3 (#4)}

\def\ARNPS{{\it Ann.\ Rev.\ Nucl.\ Part.\ Sci.\ }}
\def\IEEE{{\it IEEE Trans.\ Nucl.\ Sci.\  }}
\def\NIM{{\it Nucl.\ Instrum.\ Methods }}
\def\NIMA{{\it Nucl.\ Instrum.\ Methods} \bf A}
\def\NPB{{\it Nucl.\ Phys.} \bf B}
\def\PLB{{\it Phys.\ Lett.}  \bf B}
\def\PRL{{\it Phys.\ Rev.\ Lett.\ }}
\def\PRD{{\it Phys.\ Rev.} \bf D}

\end{document}